\documentclass[runningheads]{llncs}
	\usepackage[T1]{fontenc}
	\usepackage{float}
	\usepackage{graphicx}
\usepackage{mathrsfs}
\usepackage{cmll}
\usepackage{etex}
\usepackage{amsmath} 
\usepackage{amssymb}
\usepackage{amsfonts}
\usepackage{mathtools} 
\usepackage{stmaryrd}
\SetSymbolFont{stmry}{bold}{U}{stmry}{b}{n}
\usepackage{mathpartir}

\makeatletter
\define@key {mprset}{style}[1]{\def\TirNameStyle{#1}}
\makeatother
\mprset{style={\scriptsize\scshape}}
\usepackage{etoolbox}
\usepackage{tikz,tikz-cd}
\usetikzlibrary{arrows,arrows.meta,positioning}
\usepackage{url}
\usepackage{subcaption}
\usepackage{graphicx}
\usepackage{caption}
\usepackage[export]{adjustbox}

\usepackage{thmtools}

\newcounter{number}
\newenvironment{varitemize}
{
	\begin{list}{\labelitemii}
		{\setlength{\itemsep}{0.0mm}
			\setlength{\topsep}{0.0mm}
			\setlength{\parindent}{0.0mm}
			\setlength{\parskip}{0.0mm}
			\setlength{\parsep}{0.0mm}
			\addtolength{\leftmargin}{-\labelsep}}}
	{
	\end{list}
}

\usepackage[draft]{optional}
\usepackage{comment}
\usepackage[colorinlistoftodos,prependcaption,textsize=tiny]{todonotes}
\usepackage{xcolor}
\usepackage[all]{xy}
\usepackage{braket}
\usetikzlibrary{fit}

\usepackage{graphicx}
\graphicspath{  }

\makeatletter
\newcommand{\oset}[3][0ex]{%
	\mathrel{\mathop{#3}\limits^{
			\vbox to#1{\kern-2\ex@
				\hbox{$\scriptstyle#2$}\vss}}}}
\makeatother

\newcommand{\arrowRef}[1]{\oset{#1}{\multimap}}

\newcommand{\lis}[1]{\langle #1 \rangle} 

\newcommand{\den}[1]{\llbracket #1 \rrbracket} 
\newcommand{\sem}[1]{\llparenthesis #1 \rrparenthesis} 
\newcommand{\Q}{\mathcal{Q}}
\newcommand{\Lq}{\mathcal{L}}

\newcommand{\eval}{\mathbf{eval}} 

\newcommand{\dlambda}{\boldsymbol{\lambda}}

\newcommand{\id}{\mathrm{id}} 
\newcommand{\op}{\mathrm{op}}

\DeclareMathOperator{\One}{\mathbf{1}} 

\newcommand{\ISet}{\mathbf{ISet}}
\renewcommand{\Set}{\mathbf{Set}}
\newcommand{\PMet}{\mathbf{PMet}}
\newcommand{\Rel}{\mathbf{Rel}}

\newcommand{\Store}{\mathrm{St}}

\renewcommand{\bar}[1]{\overline{#1}}

\newcommand{\supp}[1]{\mathbf{supp}{(#1)}}
\renewcommand{\graph}[1]{\mathbf{gr}{(#1)}}

\newcommand{\BB}{\mathbb{B}}
\renewcommand{\SS}{\mathbb{S}}

\newcommand{\UU}{\mathbb{U}}

\newcommand{\N}{\mathbb{N}}
\newcommand{\Rp}{\mathbb{R}_+}

\newcommand{\lamBLL}{{\lambda}\mathbf{BLL}}

\newcommand{\Val}{\mathcal{V}}

\newcommand{\letin}[2]{\mathtt{let}\,#1\,\mathtt{in}\, #2}
\newcommand{\ifthen}[3]{\mathtt{if}\,#1\,\mathtt{then}\, #2\,\mathtt{else}\; #3}
\newcommand{\loopp}[3]{\mathtt{loop}\,#1\,#2\,\mathtt{times \, from}\, #3}
\newcommand{\setr}[2]{\mathtt{set}\,#1\, #2}
\newcommand{\return}{\mathtt{return}\,}
\newcommand{\der}{\mathtt{der}}
\newcommand{\get}{\mathtt{get}\,}

\newcommand{\true}{\mathbf{t}}
\newcommand{\false}{\mathbf{f}}

\newcommand{\typeof}[1]{\mathsf{typeof}(#1)}
\newcommand{\Fun}{\mathcal{F}}
\newcommand{\algof}{\mathsf{alg}}
\newcommand{\random}{\mathtt{random}}
\newcommand{\equals}{\mathtt{equal}}
\newcommand{\xor}{\mathtt{xor}}

\newcommand{\flipcoin}{\mathtt{flipcoin}}

\newcommand{\subst}[1]{#1}

\newcommand{\dV}{\mathbf{dV}} 
\newcommand{\dC}{\mathbf{dC}} 
\newcommand{\disc}{\mathrm{disc}} 
\newcommand{\stat}{\mathrm{stat}} 

\newcommand{\Ind}{\mathbf{Ind}} 
\newcommand{\VInd}{\mathbf{IndV}} 
\newcommand{\CInd}{\mathbf{IndC}} 
\newcommand{\TermR}{\mathcal{R}} 
\newcommand{\TermRV}{\mathcal{RV}}
\newcommand{\TermRC}{\mathcal{RC}}

\newcommand*{\smstep}[1][]{\longrightarrow_{#1}}
\newcommand*{\smsteps}[2]{\Downarrow^{#1}_{#2}}

\newcommand{\D}{\mathbf{D}} 
\newcommand{\Dist}{\mathscr{D}} 
\newcommand{\T}{\mathbf{T}}

\newcommand{\coupling}{\Omega}

\newcommand{\Barr}[1]{\overline{#1}}
\newcommand{\Kan}{\mathbf{K}}

\newcommand{\ie}{\emph{i.e.}\,}
\newcommand{\eg}{\emph{e.g.}\,}

\newcommand{\bindsymbol}{\scalebox{0.5}[1]{$>\!>=$}}
\newcommand{\bind}{\mathrel{\bindsymbol}}

\DeclarePairedDelimiter\abs{\lvert}{\rvert} 
\usetikzlibrary{decorations.markings}
\tikzset{module/.style={%
		decoration={markings,%
			mark= at position 0.5 with {%
				\node[] (tempnode) {};
				\draw[-,shorten <=.1em,shorten >=.1em,inner sep=1pt]
				(tempnode.north) -- (tempnode.south);
			}
		},
		postaction={decorate}
	}
}
\newcommand{\profto}[1][\null]{%
	\mathrel{%
		\tikz{\draw[module,->,transform canvas={yshift=3pt}%
			] (0,0) -- node[above]{$\scriptstyle #1$} ++(1/2,0) ;%
		}
	}%
}

\def\typstr#1{\mathbb{S}[#1]}
\def\typbool{\mathbb{B}}
\def\typbang#1#2{!_{#1} #2}
\def\judgment#1#2#3#4{#1\mbox{ }\mbox{ }#2 \vdash #3 : #4}
\def\Gen{\mathit{Gen}}
\def\Enc{\mathit{Enc}}
\def\Oracle{\mathit{Oracle}}
\def\Adv{\mathit{Adv}}
\def\typmsgprivk{\typstr{p_m}}
\def\typcprivk{\typstr{p_c}}
\def\typkeyprivk{\typstr{p_k}}
\def\typunit{\mathbb{U}}
\def\lollipop{\multimap}
\def\tensor{\otimes}
\def\privkcpaterm{\mathit{PrivKCPA}}
\def\privkcpa{\texttt{PrivK}^{\mathit{CPA}}}
\def\GenAlg{\mathit{Gen}}
\def\EncAlg{\mathit{Enc}}
\def\tab{\quad}
\def\typtabrand{\typstr{p_m *2n}}

\begin{document} 

\title{On Computational Indistinguishability\\ and Logical Relations}

\author{Ugo Dal Lago\inst{1,2}\orcidID{0000-0001-9200-070X} \and
			Zeinab Galal\inst{1,2}\orcidID{0009-0008-6402-3531} \and
			Giulia Giusti\inst{3}\orcidID{0000-0002-6533-8307}}
	
\institute{University of Bologna, Italy\and
	INRIA Sophia Antipolis, France \and
		ENS Lyon, France}

\maketitle 

\begin{abstract}
	A $\lambda$-calculus is introduced in which all programs can be evaluated in probabilistic 
	polynomial time and in which there is sufficient structure to represent sequential 
	cryptographic constructions and adversaries for them, even when the latter are oracle-based. 
	A notion of observational equivalence capturing computational indistinguishability 
	and a class of approximate logical relations are then presented, showing that 
	the latter represent a sound proof technique for the former. 
	The work concludes with the presentation of an example of a security proof in which 
	the encryption scheme induced by a pseudorandom function is proven secure against active 
	adversaries in a purely equational style.
\end{abstract}


\section*{Introduction}
	The two predominant models in cryptography, namely the computational~\cite{GoldwasserMicali} and the symbolic~\cite{DolevYao} models, have had very different fates with respect to the application of language-based verification techniques to them. In the symbolic model, which does not account for complexity nor for probability, the application of classic verification methodologies (e.g. model checking~\cite{fiore2001computing}, rewriting~\cite{mitchell2002multiset} and abstract interpretation~\cite{proverif}) is natural and has been extensively done. In the computational model, instead, all this is notoriously more problematic.

An interesting line of work, which has given rise to an increasing number of contributions in the last 25 years (see, e.g.,~\cite{impagliazzo_kapron,mitchell_linguistic_1998,marcinkowski_complete_2004,dal_lago_session_2022}), consists in the application of classical program equivalence theories to programming languages specifically designed to capture the reference notion of complexity in the computational model, namely that of a probabilistic polynomial time algorithm (PPT below). Once this is done, the gold standard notion of equivalence in cryptography, namely \emph{computational indistinguishability}~\cite{katz2020introduction,goldreich_foundations_2007}, becomes a form of observational equivalence, thus paving the way towards the study of computational indistinguishability via standard tools from programming language theory, like logical relations~\cite{tait1967intensional,statman1985logical} and applicative bisimilarity~\cite{abramsky1990,sangiorgi2011advanced}, which are sound by construction (although not necessarily complete) for observational equivalence.

This is precisely the direction we explore in this work; our objective is to define a typed $\lambda$-calculus with references and probabilistic choice able to naturally capture the complexity constraints mentioned above through a form of graded modality, at the same time allowing to easily express primitives, experiments and reductions, which are the building blocks on which game-based proofs are based. The language we introduce, called $\lamBLL$, can be seen as derived from Bounded Linear Logic~\cite{girard_bounded_1992}. Its syntax and operational properties are analyzed in detail in Section \ref{sec:syntax}.

In Section \ref{sec:indistinguishability}, we then move on to define a notion of logical relation for $\lamBLL$ and demonstrate that it is \emph{sound} for an approximate observational equivalence precisely capturing computational indistinguishability. A crucial aspect is that the proposed logical relation, in fact based on a logical \emph{metric}, is \emph{approximate} and therefore manages to capture programs that do not behave \emph{exactly} the same way.

Section \ref{sec:privateKey} is devoted to showing how a set of equations all justifiable through the introduced logical relations allows us to prove the security of an intrinsically second-order cryptographic construction, i.e. the proof that the encryption scheme induced by a pseudorandom function is CPA-secure, a classic result in cryptography. Notably, this proof intrinsically relies on approximate notions of equivalence. Moreover, parts of it make essential use of references.

\section{$\lamBLL$: a Calculus Capturing PPT}\label{sec:syntax}
	
We define a language $\lamBLL$, inspired by graded $\lambda$-calculi~\cite{reed_distance_2010} and CBPV~\cite{levy2012call,ehrhard2019probabilistic}, and expressive enough to model complex cryptographic experiments requiring to keep track of the messages on which the oracle is queried by the adversary. 

\subsubsection{Types}
At the level of types, $\lamBLL$ has a linear type system (Figure \ref{fig:syntax}) with a correspondence to Bounded Linear Logic ($\mathbf{BLL}$)~\cite{girard_bounded_1992} and graded-calculi~\cite{reed_distance_2010} with indexed comonadic types. In our case, the grades are polynomials and serve to keep the complexity of the attackers under control. They are built from positive natural numbers ($\N_{\geq 1}$), addition and multiplication, but also contain a polynomial variable $i$ (corresponding to the security parameter), allowing us to reason on indexed families of types and terms as in $\mathbf{BLL}$~\cite{girard_bounded_1992}. Ground types are generated from unit $\UU$, booleans $\BB$ and binary strings $\SS[p]$ of length $p$ for some polynomial $p$.  We distinguish between positive types and general types in the CBPV style~\cite{levy2012call,ehrhard2019probabilistic} to restrict the argument of an application to be of positive type. To model references, we use effect typing~\cite{gifford_integrating_1986} and annotate the bang and arrow types with reference contexts providing information on which memory cells are used during program execution. We also consider two types of contexts to distinguish between term variables $x$ and memory references $r$. 
\begin{figure}[h]
	\begin{adjustbox}{minipage=\linewidth,scale=0.84}
		\centering 
		\vspace*{-5mm}
		\begin{align*}
			&\textbf{Ground types}  && G ::= \UU \mid \BB \mid \SS[p] &\textbf{Ground values} \;& W ::= \star \mid \true \mid \false \mid s  \\
			&\textbf{Positive types} \; && P ::= G \mid P \otimes P \mid \oc_p^{\Theta} A &\textbf{Positive values} \; & Z ::= x \mid W\mid \lis{Z,Z} \mid \oc M \\
			&\textbf{Types} \; && A ::= P \mid P \arrowRef{\Theta} A &\textbf{Values}  \;  &\Val \ni V ::= Z \mid \lambda x . M\\
			& &&	&\textbf{Computations}   \; & \Lambda \ni M ::=  \return V \mid  \der (Z) \mid  MZ  \\
			&  \textbf{Variable contexts} && \Gamma ::= \varnothing \mid x :P, \Gamma & & \mid \letin{x=N}{M} \mid
			f_{p}(Z_1, \dots, Z_m)     \\
			&\textbf{Reference contexts} && \Theta ::= \varnothing \mid r : G , \Theta & & \mid   \loopp{V}{p}{M} 	\mid 	\setr{r}{Z}  \\  
			& && & &  \mid \ifthen{Z}{M}{N} \mid	 \get r \\  
			& \textbf{Polynomials} \; && p ::= 1 \mid i \mid p + p \mid p\times p 
			& &  \mid \letin{\lis{x,y}=Z}{M}   
		\end{align*}
		\vspace*{-0.5cm}
	\end{adjustbox}
	\caption{Syntax of $\lamBLL$}
	\label{fig:syntax}
\end{figure}

\subsubsection{Terms}

Grammars for values and computations are in Figure \ref{fig:syntax}. Memory references can only store values of ground type, and are handled in a simple way via reading and writing operators on locations. 
The term $\setr{r}{V}$ corresponds to updating the memory location referenced by $r$ with the value $V$ and $\get r$ returns the value under the reference $r$. 

We enrich the grammar of $\lambda$-calculus with function symbols computing probabilistic polytime functions, which are the basic building blocks of any cryptographic protocol. We fix a set of function symbols $\Fun$ and each function symbol $f$ in $\Fun$ comes equipped with:
\begin{itemize}
	\item[$\bullet$] a type denoted $\typeof{f}$ of the form $G_1 \times \dots \times G_m \to G$ where $G_1, \dots, G_m$ and $G$ are ground types;
	\item[$\bullet$] for every polynomial $p$ in  $\N_{\geq 1}[i]$, a term constructor $f_{p}$ of arity $m$.
\end{itemize}
For example, we will consider the function symbol $\random$ with $\typeof{\random} = \SS[i]$ and arity $0$ interpreted as a map randomly generating a string in $\{0,1\}^i$, and the function symbol $\xor$ with $\typeof{\xor} = \SS[i] \times \SS[i] \to \SS[i]$ and arity $2$ interpreted as a map computing the \emph{bitwise exclusive-or} of binary strings. 

Furthermore, in order to make $\lamBLL$ expressive enough to model experiments involving an adversary that can access an oracle a polynomial number of times, the grammar of computations includes an iterator $\mathtt{loop}$.

\subsubsection{Typing Rules}
The typing rules for $\lamBLL$ are given in Figure \ref{fig:typingRules}. We have two kinds of typing judgements: \[
\Gamma  \vdash V : A \quad \text{and} \quad\Gamma ; \Theta \vdash M:A
\] for values and computations respectively,
where $\Gamma = x_1 : P_1, \dots, x_n : P_n$ is a context assigning positive types to term variables and $\Theta  = r_1 : G_1, \dots, r_n : G_n$ is a reference context assigning ground types to reference variables.
The operation of polynomial addition induces a binary partial operation $\boxplus$ on positive types defined by induction below:
\[\begin{aligned}
	G \boxplus G &:= G\\
	(P \otimes Q) \boxplus(R \otimes S) &:= (P \boxplus R) \otimes (Q \boxplus S) \\
	(\oc_p^{\Theta} A) \boxplus (\oc_q^{\Theta}  A) &:= \oc_{p+q}^{\Theta}  A. 
\end{aligned}\]
To account for polynomial multiplication, we also define for every polynomial $p \in \N_{\geq 1}[i]$, a total unary operation on positive types by induction:
\[\begin{aligned}
	p\ast G &:= G\\
	p \ast (P \otimes Q)  &:= (p \ast P) \otimes (p \ast Q) \\
	p \ast (\oc_q^{\Theta}  A) &:= \oc_{p\times q}^{ \Theta}  A. 
\end{aligned}\]

It is important to note that on ground types, the identities $G \boxplus G = G$ and $p\ast G = G$ mean that ground values (unit $\star$, booleans $\true, \false$ and binary strings $s \in \{0,1\}^*$) are duplicable whereas we keep track of the polytime complexity for higher-order applications and effects similarly to~\cite{dal2012linear}.
The partial operation $\boxplus$ on positive types can be extended to a total operation on variable contexts:
\[\begin{aligned}
	\varnothing \boxplus \varnothing&:= \varnothing\\
	(x: P, \Gamma) \boxplus \Delta & := \begin{cases}
		x : P, \Gamma \boxplus \Delta & \text{ if } x \text{ does not occur in } \Delta\\
		x : P \boxplus Q, \Gamma \boxplus \Sigma & \text{ if } \Delta = x: Q, \Sigma
	\end{cases}
\end{aligned}\]
We also extend the operation $p \ast (-)$ on positive types to a total operation on term variables contexts:
\[\begin{aligned}
	p \ast \varnothing &:= \varnothing\\
	p \ast  (x : P,\Gamma )&:= (x : p \ast P),  p\ast \Gamma 
\end{aligned}\]
For a polynomial $p$ in $\N_{\geq 1}[i]$ and a type $A$, we write $A\subst{p}$ for the type $A[p/i]$ where we substitute all the occurrences of the security parameter $i$ by $p$. Similarly, for a term $M$, we write $M\subst{p}$ for the term $M[p/i]$.

\begin{figure}[h]
	\vspace*{-6mm}
	\begin{center} \scriptsize
	\begin{mathpar}\mprset{sep=1em,  vskip =-0.1ex}
		\inferrule* [right=var]{\\}{ \Gamma, x : P \vdash x: P}
		\and 
		\inferrule* [right=true]{\\}{\Gamma \vdash \true : \BB}	
		\and
		\inferrule* [right=false]{\\}{\Gamma \vdash \false : \BB}	
		\and
		\inferrule* [right=fun]{ \mathrm{typeof}(f) = G_1 \times \dots \times G_m \to G \\ (\Gamma_k\subst{p} \vdash Z_k : G_k\subst{p})_{1\leq k \leq m} \\ p \in \N_{\geq 1}[i]}{ \boxplus_k \Gamma_k\subst{p} ; \Theta \vdash f_p(Z_1, \dots, Z_m):G\subst{p}}
		\and
		\inferrule* [right=string]{s\in \{0,1\}^{c} \\ p : i \mapsto c\text{ is a constant polynomial}}{\Gamma \vdash s : \SS[p]}	
		\and
		\inferrule* [right=tensor]{\Gamma \vdash Z_1 : P \\ \Delta \vdash Z_2: Q}{\Gamma \boxplus \Delta \vdash \lis{Z_1, Z_2} : P\otimes Q}	
		\and
		\inferrule* [right=let]{\Gamma \vdash Z :  P \otimes Q\\x : P, y : Q, \Delta ; \Theta \vdash  M : A}{\Gamma \boxplus \Delta; \Theta \vdash  \letin{ \lis{x,y} = Z}{M} : A }
		\and
		\inferrule* [right=unit]{\\}{\Gamma \vdash \star : \UU}	
		\and
		\inferrule* [right=bang]{\Gamma ; \Theta \vdash M : A}{p \ast \Gamma \vdash \oc M : \oc_p^{\Theta} A}	
		\and
		\inferrule* [right=der]{\Gamma \vdash Z : \oc_1^{\Theta} A }{\Gamma;   \Theta \vdash \der (Z):A } 	
		\and
		\inferrule* [right=app]{\Gamma ; \Theta \vdash M : P\arrowRef{\Theta} A
		\\ \Delta \vdash Z : P}{\Gamma\boxplus \Delta ; \Theta \vdash MZ : A}	
		\and	\inferrule* [right=lam]{\Gamma, x: P ; \Theta \vdash M : A }{\Gamma \vdash \lambda x . M : P\arrowRef{\Theta} A}
		\and
		\inferrule* [right=eta]{\Gamma  \vdash V: A}{\Gamma ; \Theta \vdash \return V: A}	
		\and
		\inferrule* [right=let]{\Gamma ; \Theta \vdash N: P \\ x : P, \Delta ; \Theta \vdash M : A }{ \Gamma\boxplus \Delta ; \Theta \vdash \letin{ x = N}{M} : A}	
		\and
		\inferrule* [right=loop]{\Gamma \vdash V: P\arrowRef{\Theta} P \\ \Delta ; \Theta \vdash  M : P }{(p \ast \Gamma)\boxplus \Delta ; \Theta \vdash  \loopp{V}{p}{M} : P }	
		\and
		\inferrule* [right=set]{  \Gamma \vdash Z: G }{ \Gamma;  \Theta, r: G  \vdash \setr{r}{Z} : \UU }	
		\and 
		\inferrule* [right=get]{\\}{ \Gamma ; \Theta, r: G \vdash \get r : G}
		\and 
			\inferrule* [right=case]{\Gamma ; \Theta \vdash Z: \BB \\ \Delta ; \Theta \vdash  M : A \\ \Delta ; \Theta \vdash  N : A}{\Gamma \boxplus \Delta ; \Theta \vdash  \ifthen{Z}{M}{N} : A}	
	\end{mathpar}
\end{center}
	\caption{$\lamBLL$ typing rules}
	\label{fig:typingRules}
	\vspace*{-6mm}
\end{figure}

\subsubsection{Probability Distributions}
Our calculus incorporates probabilistic effects with references by combining the distribution monad and the state monad.
Recall that for a set $X$, a \emph{(finite) probability distribution} is a function $\mu : X \to [0,1]$ with finite support, \ie the set $\supp{\mu} := \{ x \in X \mid \mu(x) > 0\}$
is finite, and such that $\sum_{x\in X} \mu(x) =1$. We denote by $\delta_x : X \to [0,1]$ the \emph{Dirac distribution} mapping an element $y$ in $X$ to $1$ if $y=x$ and to $0$ otherwise.
Any probability distribution $\mu$ is then equal to 
\[
\sum_{1\leq k \leq m} a_k \cdot \delta_{x_k} \text{ where }\{x_1, \dots, x_m\} =\supp{\mu} \text{ and } a_k = \mu(x_k)
\] 
for $1\leq k \leq m$. We denote by $\D(X)$ the set of all probability distributions over $X$. It induces a monad $(\D, \eta_{\D}, \bind_{\D})$ on the category $\Set$ of sets and functions (we will omit the subscripts if there is no ambiguity). The unit has components $\eta_X : x \mapsto \delta_x$ given by Dirac distributions and the bind operator 
\[\bind \, : \D(X) \times \Set(X, \D(Y)) \to \D(Y)\] maps a distribution $\mu = \sum_k a_k \delta_{x_k} \in \D(X)$ and a function $f : X \to \D(Y)$ to the pushforward distribution $\mu \bind f := \sum_k a_k \delta_{f(a_k)}$.

\subsubsection{Combining Probability with References}\label{sec:monadT}
The general idea is that a \emph{store} is a map from memory reference variables to values that preserves typing. More precisely, for a fixed closed reference context $\Theta = r_1 : G_1, \dots, r_m : G_m$ (meaning that the security parameter variable $i$ does not occur in the types $G_1, \dots G_m$), we denote by $\Store_\Theta$ the set of functions $e : \{r_1, \dots, r_m \} \to \Val$ such that $e(r_j) \in \{ V \in \Val \mid \, \cdot \vdash V : G_j\}$ for all $1 \leq j \leq m$.

We associate to every closed $\Theta$ a corresponding monad $(\T_\Theta, \eta_\Theta, \bind_\Theta)$ on $\Set$ given by the tensor product \cite{hyland_combining_2006} of the distribution monad with the state monad $\T_{\Theta}: = (\D (- \times \Store_{\Theta}))^{\Store_{\Theta}}$, similarly to \cite{aguirre_higher-order_2021}. The unit of $\T_\Theta$ has components $X \to \D (X \times \Store_\Theta)^{\Store_\Theta}$ mapping $x \in X$ and $e \in \Store_\Theta$ to the Dirac distribution $\delta_{(x,e)}$.  The bind operator \[
\bind_\Theta : \T_\Theta X  \times \Set(X, \T_\Theta Y) \to \T_\Theta Y
\]takes $\varphi \in \T_\Theta X $ and $f : X \to \T_\Theta Y$ to the map $\dlambda e. (\varphi(e) \bind_\D \eval \circ(f \times \id_{\Store_\Theta})) $ where $\dlambda$ and $\eval$ are respectively the Currying operator and the evaluation map induced by the Cartesian closed structure of $\Set$.

\subsubsection*{From Sets to Indexed Families}
To work with general term sequents where the security parameter $i$ \emph{may occur freely}, we generalize the discussion above from sets to families of sets. Let $\ISet$ be the category whose objects are families $X = \{X_n\}_{n\geq 1}$ of sets indexed by $\N_{\geq 1}$ and a morphism from $X = \{X_n\}_{n\geq 1}$ to $Y = \{Y_n\}_{n\geq 1}$ is a family of functions $\{f_n : X_n \to Y_n\}_{n\geq 1}$.

In our calculus, probabilistic effects are generated via the function symbols in $\Fun$. For each $f \in \Fun$ with $\typeof{f} = G_1 \times \dots \times G_m \to G$, we assume that:
\begin{itemize}
	\item[$\bullet$] there is a family $\den{f} = \{ \den{f}_n \}_{n \geq 1}$ of set-functions $\den{f}_n : \den{G_1}_n \times \dots \times  \den{G_m}_n \to \D(\den{G}_n)$
	indexed over the security parameter $n \geq 1$ where $\den{\SS[p]}_n:= \{0, 1\}^{p(n)}$,  $\den{\BB}_n:= \{\true, \false\}$ and $\den{\UU}_n :=\{\star\}$.
	\item[$\bullet$] these functions can be evaluated in probabilistic polynomial time: there exists a PPT algorithm $\algof(f)$ such that for every $n\geq 1$, if $\algof(f)$ is fed with input $1^n$ and a tuple $t \in 
	\den{G_1}_n \times \dots \times  \den{G_m}_n$, it  returns $x \in \den{G}_n$ with probability $\den{f}_{n}(t)(x)$. This can be achieved by taking function symbols from a language guaranteeing the aforementioned complexity bounds \cite{mitchell_linguistic_1998,dal_lago_probabilistic_2014}.
	A very small amount of these would however be sufficient for completeness.
\end{itemize}
Now, for a general reference context $\Theta$ (whose types may contain $i$), we define a monad on $\ISet$ mapping an indexed family $X = \{X_n\}_{n\geq 1}$ to the family
\[
\{ \T_{\Theta n}(X_n)\}_{n\geq 1} =
\{
(\D (X_n \times \Store_{\Theta n}))^{\Store_{\Theta n}}
\}_{n\geq 1}
\]
which we will use for the operational semantics of $\lamBLL$.

\subsubsection{Operational Semantics}\label{sec:finalSem}
For every variable context $\Gamma$, reference context $\Theta$ and type $A$, we define indexed families $ \Lambda^\Theta(\Gamma; A) =\{ \Lambda_n^\Theta(\Gamma; A) \}_{n \geq 1}$ and $\Val(\Gamma; A) = \{\Val_n(\Gamma; A) \}_{n \geq 1}$ of typable terms and values respectively as \[\begin{aligned}
	\Lambda_n^\Theta(\Gamma; A) &:= \{ M \in \Lambda \mid  \Gamma\subst{n} ;  \Theta\subst{n} \vdash M : A\subst{n} \} \text{ and }\\
	\Val_n(\Gamma; A) & := \{ V \in \Val \mid  \Gamma\subst{n}\vdash  V : A\subst{n}\}.
\end{aligned}\]
If the variable context $\Gamma$ is empty, we write $\Lambda_n^\Theta(A)$ and $\Val_n(A)$ for $\Lambda_n^\Theta(\varnothing;A)$ and $\Val_n(\varnothing; A)$ respectively. 

For a fixed reference context $\Theta$, the small step operational semantics (Figure~\ref{fig:smallstep}) is an indexed relation $\smstep\, = \{ \smstep[n] \}_{n\geq 1}$ with
\[
\smstep[n] \, \subseteq (\Lambda^\Theta_n  \times \Store_{\Theta n}) \times \D(\Lambda^\Theta_n  \times \Store_{\Theta n})
\]
where $\Lambda^\Theta_n := \{ M \in \Lambda \mid \Gamma n ; \Theta n \vdash M : An \text{ for some } \Gamma, A \}$. For a triple $(M,e, \Dist)$ in $\smstep[n]$, we write $(M,e) \smstep[n] \Dist$ and for ease of readability, we denote a probability distribution $\sum_{1\leq k \leq m} a_k \delta_{x_k}$ in set theoretic fashion $\{ x_1^{a_1}. \dots, x_m^{a_m}\}$.
\begin{figure}[h]
	\centering
		\vspace*{-5mm}
	\begin{adjustbox}{minipage=\linewidth,scale=0.86}
		\begin{mathpar}
			\inferrule* []{\\}{ (\letin{x=\return V}{M}, e) \smstep[n] \{(M[V/x],e)^1 \} }
			\vspace{-0.2cm}
			\and
			\inferrule* []{(N, e) \smstep[n] \{(N_k, e_k)^{a_k}\} }{ (\letin{x=N}{M}, e)\smstep[n]  \{ (\letin{x=N_k}{M}, e_k)^{a_k}\} }
			\vspace{-0.3cm}
			\and
			\inferrule* []{\\}{(\letin{ \lis{x,y} = \return \lis{V,W}}{M}, e)\smstep[n] \{(M[V/x,W/y], e)^1\}}
			\and
			\inferrule* []{\\}{((\lambda x. M)V, e)\smstep[n] \{(M[V/x],e)^1\}}
			\and
			\inferrule* []{(M, e) \smstep[n] \{(M_k, e_k)^{a_k}\} }{ (MV, e)\smstep[n]  \{ (M_k V, e_k)^{a_k}\} }
			\vspace{-0.3cm}
			\and
			\inferrule* []{\\}{(\der(\oc M), e)\smstep[n] \{(M, e)^1\}}
				\and	
			\inferrule* []{\\}{( \setr{r}{V} , e)\smstep[n] \{(\star,e[V/r])^1\}}
			\vspace{-0.3cm}
			\and
			\inferrule* []{\\}{( \get{r} , e)\smstep[n] \{(e(r),e)^1\}}
			\and
			\vspace{-0.3cm}
			\inferrule* []{\\}{( \ifthen{\true}{M}{N} , e)\smstep[n] \{(M,e)^1\}}
			\and
			\inferrule* []{\\}{( \ifthen{\false}{M}{N} , e)\smstep[n] \{(N,e)^1\}}
			\vspace{-0.3cm}
			\and
			\inferrule* []{\\}{( f_p(W_1, \dots, W_m) , e)\smstep[n] \{ (\den{f}_{p(n)}(W_1, \dots, W_m), e)^1\}}
			\vspace{-0.3cm}
			\and
			\inferrule* []{\\}{( \loopp{(\lambda x. M)}{1}{N} , e)\smstep[n] \{(\letin{x=N}{M},e)^1\}}
			\vspace{-0.3cm}
			\and
			\inferrule* []{\\}{( \loopp{(\lambda x. M)}{k+1}{N} , e)\smstep[n] \{(\letin{x=(\loopp{(\lambda x.M)}{k}{N})}{M},e)^1\}}
		\end{mathpar}
	\end{adjustbox}
	\caption{Small step semantics of $\lamBLL$}
	\vspace{-0.5cm}
	\label{fig:smallstep}
\end{figure}
For the case $\setr{r}{V}$, $e[V/r]$ denotes the store mapping a reference $r'$ to $e(r')$ if $r' \neq r $ and to $V$ if $r'=r$.

Our calculus is strongly normalizing and in addition to the small step operational semantics for one step reductions, we also provide a final or big step semantics for the convergence behavior of terms. For a fixed reference context $\Theta$ and a type $A$, the final semantics $\sem{-}^{\Theta,A}$ for closed $\lamBLL$-terms is a map in $\ISet$ corresponding to the indexed family of functions
\[
\{ \sem{-}_n^{\Theta,A}: \Lambda_n^\Theta(A) \to 
\D (\Val_n(A) \times \Store_{\Theta n}))^{\Store_{\Theta n}}
\}_{n\geq 1}
\]
obtained in two steps:
\begin{enumerate}
	\item We first use the fact that the monad $\T_\Theta$ on $\Set$ extends to  the category of $\omega$-complete partial orders with a bottom element ($\omega$-cppo) and Scott-continuous morphisms (it follows easily from the fact that both the distribution and the state monads extend to $\omega$-cppo's). It allows us to define inductively a family $\sem{-}_n^{\Theta, A}: \Lambda_n(A) \to \T_{\Theta n}^\bot(\Val_n(A))$ where for a set $X$,  $\T_{\Theta n}^\bot:= \T_{\Theta n}(X \uplus \{ \bot\}, \leq)$ is the image of the flat ordering ($\bot \leq x$ for all $x \in X$) under $\T_{\Theta n}$ (the bottom element $\bot$ is added to account for computations which are possibly non-terminating). Similarly to \cite{leroy_coinductive_2009}, each map $\sem{-}_n^{\Theta, A}$ is obtained as the supremum $\bigvee_{k\in \omega}	\sem{-}_{n,k}^{\Theta, A}$ where  $\sem{-}_{n,k}^{\Theta, A}$ is defined inductively below:
	\begin{center}
		\begin{adjustbox}{minipage=\linewidth,scale=0.9}
			\centering
			\begin{align*}
				&\sem{M}^{\Theta, A}_{n,0} := \bot &&\sem{\return {V} }^{\Theta,A}_{n,k+1}:= \eta_{\Val_n(A)}(V)\\
				&\sem{\ifthen{\true}{M}{N}}^{\Theta, A}_{n,k+1} := \sem{M}^{\Theta, A}_{n,k}&&
				\sem{\ifthen{\false}{M}{N}}^{\Theta, A}_{n,k+1} := \sem{N}^{\Theta, A}_{n,k}\\
				&\sem{\setr{r}{Z}}^{\Theta, A}_{n,k+1}(e) := \delta_{(\star, e[Z/r])}&& \sem{\get{r}}^{\Theta, A}_{n,k+1}(e) :=	\delta_{(e(r),e)}\\
			\end{align*}
			\vspace{-1.4cm}
			\begin{align*}
				\sem{\der (\oc M)}^{\Theta, A}_{n,k+1}& := \sem{M}^{\Theta, A}_{n,k} \\
				\sem{\letin{x=N}{M}}^{\Theta, A}_{n,k+1} &:= \sem{N}^{\Theta, P}_{n,k} \bind (U \mapsto \sem{M[U/x]}^{\Theta, A}_{n,k})\\
				\sem{\letin{ \lis{x,y} = \lis{Z,Z'}}{M}}^{\Theta, A}_{n,k+1} &:=  \sem{M[Z/x, Z'/y]}^{\Theta, A}_{n,k} \\
				\sem{MZ}^{\Theta, A}_{n,k+1} &:= \sem{M}^{\Theta, P \arrowRef{\Theta} A}_{n,k} \bind (\lambda x. N \mapsto \sem{N[Z/x]}^{\Theta, A}_{n,k})\\
			\end{align*}
		\end{adjustbox}
	\end{center}
For the case $\sem{\loopp{\lambda x. M}{m}{N}}^{\Theta, A}_{n,k+1}$, if $m=1$, we define it to be $\sem{N}^{\Theta, A}_{n,k}  \bind (U \mapsto \sem{M[U/x]}^{\Theta, A}_{n,k} )$ and if $m> 1$, we take 
\[
\sem{\loopp{\lambda x. M}{(m-1)}{N}}^{\Theta, A}_{n,k} \bind (U \mapsto \sem{M[U/x]}^{\Theta, A}_{n,k} ).
\]
For a function symbol $f$ with $\typeof f = G_1 \times \dots \times G_m \to G$ and a polynomial $p$, $\sem{f_p (W_1, \dots, W_m)}^{\Theta, A}_{n,k+1}(e)$ is the mapping 
\[
(V,e')\mapsto \delta_e(e')\den{f}_{p(n)}(W_1, \dots, W_m)(V).
\]
	\item We prove that for any $M \in \Lambda_n^\Theta(A)$ and $e \in \Store_{\Theta n}$, $(M,e)$ reduces to some distribution $\Dist$ in polynomial time. As a corollary, we obtain that the final semantics map $\sem{-}_n^{\Theta, A}$ can in fact be restricted to $\Lambda^\Theta_n(A) \to \T_{\Theta n}(\Val_n(A))$ since $\lamBLL$ is strongly normalizing.
\end{enumerate}
To express the standard correspondence between the final (big step) semantics and the transitive closure of the small step semantics, we define an indexed relation 
\[\Downarrow \,= \{ \smsteps{m}{n}\,
\subseteq (\Lambda^\Theta_n  \times \Store_{\Theta n}) \times \D(\Lambda^\Theta_n  \times \Store_{\Theta n}) \}_{n\geq 1, m \geq 0}\]
where the additional natural number $m$ models the number of steps in the small step semantics:
\begin{mathpar}\mprset{sep=1em,  vskip =0.2ex}
	\inferrule*[]{\\}{(\return V,e) \smsteps{0}{n} \{(\return V,e)^1 \}}
	\and
	\inferrule*[] {(M,e) \smstep[n] \Dist \\ \{E \smsteps{m_k}{n} \mathscr{E}_E\}_{ E\in\supp{\Dist}}}{(M,e) \smsteps{1+ \max_k m_k}{n} \textstyle \sum_E \Dist(E) \cdot \mathscr{E}_E }
\end{mathpar}
and formulate the result as follows:
\begin{restatable}{lemma}{finalSmallStep}
	\label{lem:finalSmallStep}
	For a fixed reference context $\Theta$, type $A$ and security parameter $n$, the following are equivalent: for any term $M \in \Lambda_n^\Theta(A)$, store $e \in \Store_{\Theta n}$ and distribution $\Dist \in\D (\Val_n(A) \times \Store_{\Theta n})$: 
	\[
	\sem{M}_n^{\Theta, A}(e) =\Dist \quad  \Leftrightarrow \quad \exists k \in \N, \;\sem{M}_{n,k}^{\Theta, A}(e) =\Dist \quad \Leftrightarrow \quad \exists m \in \N, (M,e) \smsteps{m}{n} \Dist
\]
\end{restatable}

\subsubsection{Soundness and Completeness for Polynomial Time}\label{sec:polytimesoundness}

A calculus like $\lamBLL$ makes sense, particularly in view of the cryptographic applications that we will present in the last part of this article, if there is a correspondence with the concept of probabilistic polynomial time. This section is dedicated to giving evidence that such a correspondence indeed holds.

Before moving on to the description of soundness and completeness, however, it is worth outlining what is meant in this context by probabilistic polynomial time. In fact, what we mean by a PPT function can be deduced from how we defined function symbols in $\Fun$: these are families of functions, indexed on natural numbers, which possibly return a distribution and are computable by a probabilistic Turing machine working in polynomial time on the value of the underlying parameter. That basic functions are PPT holds by hypothesis, but that the same remains true for any term definable in the calculus needs to be proved. Moreover, the fact that any such function can be represented in $\lamBLL$ has to be proved as well.

\paragraph{Soundness for PPT}
The goal is to show that there exists a polynomial bound on the length of reduction sequences for any term of $\lamBLL$:
\begin{theorem}[Polytime Soundness]\label{th:polysound}
    For every type derivation $\pi$ of a term $M$ in $\lamBLL$, there exists a polynomial $q_\pi$ such that for every natural numbers $n \geq 1,m \geq 0$ and store $e$, if $(M\subst{n}, e) \smsteps{m}{n}\Dist$, then $m\leq q_\pi(n)$.
\end{theorem}
Similarly to Section 4.2 in~\cite{dal_lago_session_2022}, the proof of Theorem \ref{th:polysound} 
is structured into three steps:
\begin{itemize}
    \item[$\bullet$] We assign a polynomial $q_\pi$ to every type derivation $\pi$ defined by induction on the structure of $\pi$.
    \item[$\bullet$] We prove that $q_{(\cdot)}$ is stable under
    polynomial substitution: for every type 
    derivation $\pi$ with conclusion $\Gamma ; \Theta \vdash M: A$ and for every 
    polynomial $p$, there is a type derivation 
    $\zeta$ with conclusion $\Gamma\subst{p} ; \Theta\subst{p} \vdash 
    M\subst{p}: A\subst{p}$ such that $q_\zeta(n) = q_\pi(p(n))$ for all $n\geq 1$.
     \item[$\bullet$] Finally, we prove that $q_{(\cdot)}$ strictly decreases along 
    term reduction:  if $\pi$ derives $N$ and $(N,e) 
    \smstep[n] \Dist$, then for all $(N',e')$ in $\supp{\Dist}$, there exists a type derivation 
    $\zeta$ for $N'$ such that $q_\pi> q_\zeta $.
\end{itemize}

\paragraph{Completeness for PPT}

Theorem~\ref{th:polysound} implicitly tells us that algorithms formulated as
typable $\lamBLL$ terms are PPT, since the number of reduction steps performed 
is polynomially bounded and reduction can be simulated by a Turing machine~\cite{lmcs:1213,lmcs:1627}. 
One can further prove that all PPT functions can be represented by 
$\lamBLL$ terms. Given the freedom we have about picking more and more basic 
function symbols, this does not seem surprising: we are anyway allowed to throw in
new basic function symbols whenever needed. However, one can prove that 
completeness for PPT can be achieved with a very minimal set of basic functions 
symbols only including cyclic shift functions on strings and functions 
testing the value of the first bit in a string.
Noticeably, soundness and completeness as presented above scale to 
\emph{second-order} PPT, namely a notion of probabilistic polynomial time 
function accessing an oracle. This is quite relevant in our setting, given our 
emphasis on cryptographic constructions, and the fact that adversaries for some 
of those have oracle access to the underlying primitive.

\section{Computational Indistinguishability}\label{sec:indistinguishability}
	
In this section, we define logical relations which we show to be sound for computational indistinguishability and which we will use in Section \ref{sec:privateKey} to prove security of the private key encryption scheme induced by a pseudorandom function. Our approach is to first define a \emph{logical metric} on terms from which we derive the indistinguishability logical relation containing terms whose distance is negligible with respect to this metric.

\subsubsection{Term Relations}

In Section \ref{sec:finalSem}, we have considered indexed families $ \Lambda^\Theta(\Gamma; A)$ and $\Val(\Gamma; A)$ of sets containing terms that are closed for the security parameter variable $i$. On the other hand, computational indistinguishability, which is the main focus of this paper, is a relation between terms where the security parameter is a free variable and can be instantiated for every positive natural number $n$.

For a variable context $\Gamma$, a location context $\Theta$ and a type $A$, we let $\Lambda_o^\Theta(\Gamma; A)$ and $\Val_o(\Gamma; A) $ be the sets of derivable computation terms and values respectively which are open for the security parameter variable $i$:
\[
\Lambda_o^\Theta(\Gamma; A) := \{ M  \in \Lambda \mid  \Gamma ; \Theta  \vdash M : A \} \qquad \Val_o(\Gamma; A) := \{ V \in \Val \mid  \Gamma \vdash  V : A\}.
\]

\begin{restatable}{lemma}{polyvarsubst}
	\label{lem:polyvarsubst}
	The following rules are derivable for all $n\geq 1$:
	\begin{mathpar}\mprset {sep=2em}
		\inferrule* []{ \Gamma ; \Theta \vdash M: A
		}{\Gamma\subst{n} ; \Theta\subst{n}  \vdash M\subst{n} : A\subst{n} }	
	\and 
	\inferrule* []{ \Gamma\vdash V: A
	}{\Gamma\subst{n}   \vdash V\subst{n} : A\subst{n} }	
	\end{mathpar}
	It implies that if $M$ is in $\Lambda_o^\Theta(\Gamma; A)$, then $M\subst{n} $ is in $\Lambda_n^\Theta(\Gamma; A)$ for all $ n\geq 1$ and a similar statement holds for values in $\Val_o(\Gamma; A)$.
\end{restatable}

\begin{definition}\label{def:TermRel}
	An \emph{open term relation} $\TermR$ is an indexed family of pairs of relations $ \{ (\TermRC^\Theta(\Gamma; A), \TermRV(\Gamma; A)) \}_{\Gamma, \Theta, A}$ with
	\[
	\TermRC^\Theta(\Gamma; A) \subseteq \Lambda_o^\Theta(\Gamma; A) \times \Lambda_o^\Theta(\Gamma; A) \qquad 	\TermRV(\Gamma; A) \subseteq \Val_o(\Gamma; A) \times \Val_o(\Gamma; A).
	\]
	A \emph{closed} (for term variables) \emph{term relation} $\TermR$ is an indexed family of pairs of relations $ \{ (\TermRC^\Theta(A), \TermRV(A)) \}_{ \Theta, A}$ with $\TermRC^\Theta(A) \subseteq \Lambda_o^\Theta(A) \times \Lambda_o^\Theta(A)$ and $\TermRV(A) \subseteq \Val_o(A) \times \Val_o(A)$.
\end{definition}

Every open term relation induces a closed term relation by restricting to empty term variable contexts. For the other direction, we use the standard notion of \emph{open extension} of a closed relation via substitutions with positive values.

\subsubsection{Contextual Indistinguishability}\label{sec:contextIndis}
The notion of behavioral equivalence we consider here is \emph{computational indistinguishability} with respect to a polytime adversary represented as a $\lamBLL$-context.
Recall that a function which grows asymptotically slower than the inverse of any polynomial is called negligible \cite{bellare_note_2002}:
\begin{definition}\label{def:negl}
	A function $\varepsilon : \N \to \Rp$ is \emph{negligible} if for all $k \in \N$, there exists $N \in \N$ such that for all $n \geq N$, $\varepsilon(n) < \frac{1}{n^k}$.
\end{definition}

\begin{definition}
	For terms $M,N$ in $\Lambda_o^\Theta(\Gamma;A)$, we say that $M$ and $N$ are \emph{contextually indistinguishable} if for every closing context $C$ such that $C[M]$ and $C[N]$ are in $\Lambda_o^{\Xi}( \BB)$ for some reference context $\Xi$, there exists a negligible function $\varepsilon : \N \to \Rp$ such that for every $n \geq 1$, $e \in \Store_{\Xi n}$ and subset $X \subseteq \{\true, \false\} \times \Store_{\Xi n}$,
	\[
	\abs{ \sem{C[M]\subst{n}}_n^{\Xi, \BB} (e)(X) - \sem{C[N]\subst{n}}_n^{\Xi, \BB} (e)(X)} \leq \varepsilon(n).
	\]
\end{definition}

We adopt a coinductive characterization of contextual indistinguishability following the approach in~\cite{gordon1998operational,lassen1998relational} in the case of contextual equivalence for applicative bisimilarity. The contextual indistinguishability relation can indeed be alternatively defined as the largest open $\lamBLL$-term relation that is both \emph{compatible} and \emph{adequate}. Compatibility means that the relation is closed under contexts: if $(M,N)$ is in $\TermR$ and $C$ is a context, then $(C[M], C[N])$ is also in $\TermR$. Adequacy on the other hand depends on the observational behavior we consider, it is typically termination for contextual equivalence or probability of convergence for non-deterministic calculi. In our case, the notion of interest in computational indistinguishability:
\begin{definition}
	We define an indexed relation $\approx\: = \{\approx_\Theta \; \subseteq \Lambda_o^\Theta(\BB)  \times \Lambda_o^\Theta(\BB)\}_\Theta$ on closed (for term variables) terms of Boolean type which are indistinguishable: a pair $(M,N)$ is in the relation $\approx_\Theta $ if and only if there exists a negligible function $\varepsilon : \N \to \Rp$ such that for every $n \geq 1$, $e \in \Store_{\Theta n}$ and subset $X \subseteq \{\true, \false\} \times \Store_{\Theta n}$,
	\[
	 \abs{ \sem{M\subst{n}}_n^{\Theta, \BB} (e)(X) - \sem{N\subst{n}}_n^{\Theta, \BB} (e)(X)} \leq \varepsilon(n).
	\]
\end{definition}

We call an open $\lamBLL$-term relation $\TermR$ \emph{adequate} if it is included in $\approx$ for closed Boolean terms, \ie $\TermR^{\Theta}(\BB) \subseteq \, \approx_\Theta$ for all reference contexts $\Theta$ and we obtain that the predicate of adequacy on open $\lamBLL$-term relations is closed under countable unions and relational composition.

\begin{restatable}{lemma}{ctxtLargestAdeqComp}\label{lem:ctxtLargestAdeqComp}
	Contextual indistinguishability is the largest adequate compatible open $\lamBLL$-relation and we denote it by $\sim$.
\end{restatable}
This coinductive characterization provides a useful proof principle to show soundness: any open relation $\TermR$ that is both compatible and adequate must be included in $\sim$ and is therefore \emph{sound} for contextual indistinguishability (\ie any pair of terms $(M,N)$ in $\TermR$ are contextually indistinguishable).

\subsection*{Background on Metrics}\label{sec:logicalMetric}
	
\subsubsection{Weighted Relations}\label{subsec:weightedRel}
For metric reasoning on $\lamBLL$, we consider distances valued in the unit real interval $[0,1]$ equipped with the operation of \emph{truncated addition} $x \oplus y := \min\{1, x+y\}$ for $x,y \in [0,1]$. We have in particular that $1 =1\oplus 1$ which has a direct correspondence with the fact that values of ground type are arbitrarily duplicable in $\lamBLL$ (see Remark \ref{rmk:truncatedSum}).

Recall that a (unital) \emph{quantale} is a tuple $(\Q, \leq, \otimes, 1)$ where $(\Q, \leq)$ is a complete lattice, $(\Q, \otimes, 1)$ is a monoid and $\otimes$ distributes over arbitrary joins \cite{rosenthal_quantales_1990}. The unit interval with the opposite of the natural order (the natural order is defined as: $x \leq y$ if and only if there exists $z$ such that $x \oplus z =y$) can be equipped with a quantale structure $\Lq =([0,1], \geq, \oplus, 0)$, called the \emph{{\L}ukasiewicz quantale}.

For sets $X$ and $Y$, an \emph{$\Lq$-weighted relation} $R : X \profto Y$ from $X$ to $Y$ consists of a function $X \times Y \to [0,1]$. They form a category, which we denote by $\Rel_{\Lq}$, where the identity $\id_X : X \profto X$ maps a pair $(x,y)$ to $0$ if $x=y$ and $1$ otherwise. The composite of two relations $R : X \profto Y$ and $S : Y \profto Z$ is the relation $S \circ R : X \profto Z$ mapping a pair $(x,z)$ to $ \inf_{y \in Y} R(x,y) \oplus S(y, z)$.
The category $\Rel_{\Lq}$ can be equipped with a \emph{dual} (transpose) operation mapping a relation $R : X \profto Y$ to the relation $R^\op : Y \profto X$ which simply maps $(y,x)$ to $R(x,y)$.
Any function $f : X \to Y$, induces a $\Lq$-relation via its graph $\graph{f} : X \profto Y$ mapping a pair $(x,y)$ to $0$ if $f(x)=y$ and $1$ otherwise. 
\subsubsection{Pseudo-metric Spaces}
If we restrict to the special case of weighted endo-relations $R: X \profto X$ that are reflexive ($R \leq \id_X$), symmetric ($R = R^\op$) and transitive $(R \leq R \circ R$), we obtain the notion of pseudo-metric space:

\begin{definition}
	A \emph{pseudo-metric space} consists of a pair $(X, d_X)$ where $X$ is a set and $d_X$ is a function from $X \times X\to [0,1]$ satisfying the following axioms:
	\begin{varitemize}
		\item reflexivity: for all $x$ in $X$, $d_X(x,x) =0$;
		\item symmetry: for all $x,y$ in $X$, $d_X(x,y) = d(y,x)$;
		\item triangular inequality: for all $x,y,z$ in $X$, $d_X (x,z) \leq d_X(x,y) \oplus d_X(y,z)$
	\end{varitemize}
	If $d_X$ further satisfies the separation axiom (for all $x,y$, $d_X(x,y)=0$ implies $x=y$), then $(X,d_X)$ is a \emph{metric space}. 
\end{definition}
For the rest of the paper, even if we do not assume that the separation axiom holds, we will just say metric space instead of pseudo-metric space.
Note that metric space with the \emph{discrete metric} $\disc : X \times X \to [0,1]$ mapping a pair $(x,y)$ to $0$ if $x=y$ and $1$ otherwise corresponds exactly to the identity weighted relation defined above.

\begin{definition}
		For two metric spaces $(X, d_X)$ and $(Y, d_Y)$, a function $f : X \to Y$ is said to be \emph{non-expansive} if for all $x,x'$ in $X$, $d_Y(f(x), f(x')) \leq d_X(x,x')$.
	We denote by $\PMet$ the category of pseudo-metric spaces and non-expansive maps.
\end{definition}

The category $\PMet$ is equivalent to the category of $\Lq$-enriched categories and $\Lq$-enriched functors between them \cite{hofmann_monoidal_2014}. We recall below some properties of $\PMet$ which we will use to define the logical metric in the following section, they are all instances of more general statements on quantale-enriched categories and we refer the reader to \cite{hofmann_monoidal_2014} for a complete account. 
$\PMet$ is symmetric monoidal closed with tensor product $(X, d_X) \otimes (Y d_Y)$ given by $(X \times Y, d_{X\otimes Y})$ where for all $x,x' \in X$ and $y,y' \in Y$, \[
d_{X\otimes Y}((x,y), (x',y')) := d_X(x, x') \oplus d_Y(y,y').
\]
The unit is given by $\One = (\{\star\}, \disc)$ and the linear hom $X \multimap Y$ has underlying set $\PMet(X,Y)$ (the set of non-expansive maps from $X$ to $Y$) and distance $d_{X\multimap Y}(f,g) :=  \sup_{x \in X} d_Y(f(x),g(x))$.
For any $k\geq 1$ and $x \in [0,1]$, we define inductively $k \cdot x$ as $1\cdot x:=x$ and $(k+1)\cdot x := (k \cdot x) \oplus x$. This operation induces a \emph{scaling} operation $k \cdot (X, d_X):=(X, k \cdot d_{X})$ on $\PMet$ which we use to model the graded bang of $\lamBLL$. Note that if $f :(X, d_X) \to (Y, d_Y)$ is non-expansive, then $f: (X, k \cdot d_X) \to (Y, k\cdot d_Y)$ is also non-expansive for all $k\in \N$.

\subsubsection{Extending Monadic Effects from Sets to Metric Spaces}
\label{subsec:laxExtensions}

In order to define the logical metric on computation terms, we need to extend the effect monad $\T_\Theta$ defined in Section \ref{sec:monadT} from sets to metric spaces.
To do so, we follow the standard approach of monad extensions from sets to quantale weighted relations~\cite{hofmann_monoidal_2014,balan2019extending}. It is well-known that the distribution monad $\D$ (and therefore the monads $\T_\Theta$ as well) on sets only \emph{laxly} extends to weighted relations via \emph{Kantorovich lifting}~\cite{kantorovich_translocation_2006,van2005metric,baldan2018coalgebraic}. The Kantorovich lifting for distributions fits into the more general framework of \emph{Barr extensions} for monads from sets to quantale relations \cite{hofmann_monoidal_2014}. 

In this section, we only give the explicit definition of how the Barr lax extension $\Barr{\T}_{\Theta}$ of the effect monad acts on metric spaces and we refer the reader to \cite{gavazzo2018quantitative,wild2022characteristic} for more background on lax extensions for weighted relations. The Kantorovich lifting can be formulated in terms of couplings for probability distributions:

\begin{definition}
	For sets $X,Y$ and distributions $\mu \in \D(X)$, $\psi \in \D(Y)$, a \emph{coupling over $\mu$ and $\nu$} is a distribution $\gamma\in \D(X \times Y)$ such that
	\[
		\forall x\in X, \mu(x) = \sum\limits_{y\in Y} \gamma(x,y) \text{ and } \forall y\in Y, \nu(y) =\sum\limits_{x\in X} \gamma(x,y).
	\]
	We denote by $\coupling(\mu, \nu)$ the set of all couplings over $\mu$ and $\nu$.
\end{definition} 
For a metric space $(X, d_X)$, the Kantorovich lifting of the distance $d_X$ is the distance $\Kan(d_X)$ on $\D(X)$ mapping distributions $\mu,\nu \in \D(X)$ to
\[
\Kan(d_X)(\mu,\nu) := \inf\limits_{\gamma \in \coupling(\mu, \nu)} \sum\limits_{x_1, x_2\in X} \gamma(x_1, x_2)\cdot d_X(\mu(x_1), \nu(x_2)).
\]
While there are many other possible choices of metrics on distribution spaces besides the Kantorovich distance $\Kan(d_X)$ \cite{gibbs_choosing_2002}, it is the smallest among the ones which laxly extends to weighted relations and it also coincides with the \emph{statistical distance} (or \emph{total variation distance}) when $d_X$ is the discrete metric.

\begin{definition}\label{def:statDistance}
	For a set $X$, the \emph{statistical distance} $d_{\stat} : \D(X) \times \D(X) \to [0,1]$ maps two distributions $\mu, \nu \in \D(X) $ to 
	\[
	d_{\stat}(\mu, \nu):=\frac{1}{2} \cdot \sum_{x\in X} \abs{\mu (x) - \nu(x)} = \sup_{A \subseteq X}\abs{\mu(A) - \nu(A)}.
	\]
\end{definition}

For a closed (for the security parameter variable) location context $\Theta$, the action of the lax extension  $\Barr{\T}_{\Theta}$ on a metric space $(X, d_X)$ is the metric space with underlying set $\T_{\Theta}(X)$ and distance $\Barr{\T}_{\Theta}(d_X)$ mapping functions $\varphi, \psi : \Store_{\Theta} \to \D (X \times \Store_{\Theta})$ to
\begin{equation}
\Barr{\T}_{\Theta}(d_X)(\varphi, \psi):= \sup_{e\in \Store_{\Theta}} \Kan(d_{X \otimes \Store}) (\varphi(e), \psi(e)).
\label{eqn:laxExtMonad}
\end{equation}

\subsubsection{Logical Metric}\label{sec:logicalMet}
We now have all the ingredients to define a logical metric for $\lamBLL$-terms using the lax extension of the monad $\T_{\Theta}$ to metric spaces.
\begin{definition}
	We define a family of metrics on closed computations and values indexed by the security parameter:
	\[
	\dV^{A}_{n} : \;  \Val_n(A) \times \Val_n(A) \to [0,1] \quad \text{and} \quad
	\dC^{\Theta, A}_{n} : \;  \Lambda_n^\Theta(A) \times \Lambda_n^\Theta(A)  \to [0,1]
	\]
	by mutual induction on the type $A$:
	\vspace{-0.5cm}
	\begin{center}
	\begin{adjustbox}{minipage=\linewidth}
		\centering
	\begin{align*}
		&\dV^{ \SS[p]}_{n} (s,s') := \disc_{ \SS[p(n)]}(s,s') && 
		 \dV^{ \BB}_{n}(W,W') := \disc_{\BB}(W,W')\\
		& \dV^{ \UU}_{n} (\star, \star) :=\disc_{\UU}(\star, \star)= 0 &&
	\dV^{\oc_p^{\Theta} A}_{n}(\oc M, \oc N) := \oplus_{p (n)} \dC^{\Theta, A}_{n}( M,  N) 
\end{align*}
\vspace{-0.8cm}
	\begin{align*}
	\dV^{P \otimes Q}_{n} (\lis{U,V}, \lis{U',V'}) &:=  \dV^{P}_{n}(U,U')\oplus  \dV^{Q}_{n}(V,V') \\
	\dV^{P \arrowRef{\Theta} A}_{n}(\lambda x. M, \lambda y. N)& := 
	 \sup_{V\in \Val_n(P)} \dC^{\Theta,A}_{n}(M[V/x], N[V/y]) \\	\dC^{\Theta,A}_{n}(M,N) &:= \bar{\T}_{\Theta}(	\dV^{A}_{n}) (\sem{M}_n^{\Theta, A}, \sem{N}_n^{\Theta, A})
	\end{align*}
\end{adjustbox}
\end{center}
where $\disc$ denotes the discrete metric and $\bar{\T}_\Theta$ is the lax extension of the functor $\T_\Theta : \Set \to \Set$ defined in (\ref{eqn:laxExtMonad}).
\end{definition}

In our setting, the metric version of the fundamental lemma states that substitution by positive value terms is a non-expansive operation:
\begin{restatable}[Fundamental Lemma for Logical Metrics]{lemma}{metricFundamentalLemma}\label{lem:metricFundamentalLemma}
	For a context $\Gamma =  x_1 : P_1, \dots, x_m : P_m$ and a term $M$ in $\Lambda_n^{\Theta}(\Gamma; A)$ with $n\geq 1$, for every closed positive values $Z_j, Z_j'  \in \Val_n(P_j)$ with $1\leq j \leq m$, we have
	\[
	\dC^{\Theta, A}_{n}(M\rho, M\rho') \leq \bigoplus\limits_{1\leq j\leq m} \dV^{P_j}_{n}(Z_j, Z_j')
	\]
	where $\rho := [Z_1/x_1, \dots, Z_m/x_m]$ and $\rho' := [Z_1'/x_1, \dots, Z_m'/x_m]$. A similar statement holds for open value terms in $\Val_n(\Gamma; A)$.
\end{restatable}

\begin{remark}\label{rmk:truncatedSum}
	A key ingredient in the proof of the fundamental lemma is the equality $\dV^{P \boxplus Q}_{n}(V,W) =	\dV^{P}_{n}(V,W)  \oplus	\dV^{ Q}_{n}(V,W)$ for closed values $V, W$, it allows to keep a precise track of how distances are amplified when contexts are added $\Gamma \boxplus \Delta$ in rules such as $\mathtt{let}$ or $\otimes$ for example. We can see here that the main motivation behind using truncated addition $\oplus$ in our setting is that it allows for the additional flexibility of having ground types being duplicable without loosing the ability to measure distances for higher types: if $P$ and $Q$ are equal to some ground type $G$, and $V \neq W$, then the equality above indeed rewrites to $1 = 1\oplus 1$ which would not be possible if we had considered for example the Lawvere quantale with regular addition instead of the {\L}ukasiewicz quantale with truncated addition.
\end{remark}

\subsubsection{Indistinguishability Logical Relation}
We now define a closed (for term variables) $\lamBLL$-relation $\Ind = (\CInd, \VInd)$ with 
\[
\CInd^{\Theta}(A)\; \subseteq \;  \Lambda_o^\Theta(A) \times \Lambda_o^\Theta(A)\quad \text{and} \quad \VInd(A)\; \subseteq \;  \Val_o(A) \times \Val_o( A).
\]
For terms $M,N$ in $\Lambda_o^\Theta(A)$, the pair $(M,N)$ is in $\CInd^{\Theta}(A)$ if there exists a negligible function $\varepsilon : \N \to \Rp$ such that for all $ n \geq  1$,
\[
\dC^{\Theta,A}_{n}(M\subst{n},N\subst{n}) \leq \varepsilon(n).
\] 
The relation on values $\VInd(A)$ is defined similarly via the logical metric on values $\dV^A$.

The fundamental lemma for the indistinguishability logical relation can now be directly derivable from the non-expansiveness of the logical metric (Lemma~\ref{lem:metricFundamentalLemma}) and basic closure properties of negligible functions:
\begin{restatable}{lemma}{fundLemmaIndis}\label{lem:fundLemmaIndis}
	For a variable context $\Gamma = x_1: P_1, \dots, x_m: P_m$ and closed positive values $(Z_k, Z_k')$ in $\VInd(P_k)$ with $1 \leq k \leq m$, we have for all $ M \in \Lambda_o^\Theta(\Gamma; A)$ and $U \in \Val_o(\Gamma; A)$, 
	\[(M\rho, M\rho') \in \CInd^{\Theta}(A) \text{ and } (U\rho, U\rho') \in \VInd(A)\]
	where $\rho:=[Z_1/x_1, \dots, Z_m/x_m]$ and $\rho' :=[Z_1'/x_1, \dots, Z_m'/x_m]$. In particular, for a closed term $M \in \Lambda^\Theta(A)$, we have $(M,M) \in \CInd^{\Theta}(A)$.
\end{restatable}

\begin{restatable}{theorem}{soundness}\label{thm:soundnessContextInd}
	The open extension of $\Ind$ is adequate and compatible.
\end{restatable}

Since the contextual indistinguishability relation $\sim$ is the largest compatible adequate relation, it contains $\Ind$ by Lemma \ref{lem:ctxtLargestAdeqComp}, which implies that $\Ind$ is \emph{sound} for contextual indistinguishability. Full abstraction on the other hand is not possible within our framework: since base types are equipped with the discrete metric whose Kantorovich lifting coincides with statistical distance, we cannot hope to capture the whole contextual indistinguishability relation as it is well-known that statistical closeness is strictly included in computational indistinguishability (\eg Proposition $3.2.3$ in \cite{goldreich_foundations_2007}).

\section{Proving Encryption Scheme Secure Equationally}\label{sec:privateKey}
	
\def\typstr#1{\mathbb{S}[#1]}
\def\typbool{\mathbb{B}}
\def\typbang#1#2{!_{#1} #2}
\def\judgment#1#2#3#4{#1\mbox{ }\mbox{ }#2 \vdash #3 : #4}
\def\Gen{\mathit{Gen}}
\def\Enc{\mathit{Enc}}
\def\Oracle{\mathit{Oracle}}
\def\Adv{\mathit{Adv}}
\def\typmsgprivk{\typstr{p_m}}
\def\typcprivk{\typstr{p_c}}
\def\typkeyprivk{\typstr{p_k}}
\def\typunit{\mathbb{U}}
\def\lollipop{\multimap}
\def\tensor{\otimes}
\def\privkcpaterm{\mathit{PrivKCPA}}
\def\privkcpa{\texttt{PrivK}^{\mathit{CPA}}}
\def\GenAlg{\mathit{Gen}}
\def\EncAlg{\mathit{Enc}}
\def\DecAlg{\mathit{Dec}}
\def\tab{\quad}
\def\typtabrand{\typstr{p_m *2n}}
\def\coarseq{\asymp}

This section is devoted to the presentation of a game-based proof~\cite{Shoup2004}
of security against active attacks for the encryption scheme 
$\Pi_F$ induced by any pseudorandom function $F$. The proof is rather standard and
a less formal version of it can be found in many cryptography textbooks
(see, \eg, \cite{katz2020introduction}). Following the advice of the anonymous
reviewers, we are keeping the presentation as self-contained as possible.

\subsubsection*{Pseudorandom Functions and Private-key Encryption Schemes}
A \emph{pseudorandom function}~\cite{katz2020introduction} is a function computed by any deterministic polytime algorithm taking two strings 
in input, and producing a string as output, in such a way that when the first of 
the two parameters is picked at random, the unary function obtained through
currying is indistinguishable from a random one, all this to the eyes of adversaries working
in probabilistic polynomial time. The notion of a pseudorandom function is closely
related to that of a secure block-cipher.

Private key encryption schemes~\cite{katz2020introduction}, instead, are triples of algorithms in the form $(\GenAlg,\EncAlg,\DecAlg)$,
where $\mathit{Gen}$ is responsible for generating a private key at random,
$\mathit{Enc}$ is responsible for turning a message into a ciphertext and $\mathit{Dec}$ is responsible for turning a ciphertext into a message. Both $\mathit{Enc}$ and $\mathit{Dec}$ make essential use of a
private shared key. One way to construct private key encryption schemes is by
way of pseudorandom functions: given one such function $F$, the 
scheme $\Pi_F$ is such that $\EncAlg$ encrypts a message $m$ as the pair $(r, F_k(r)\oplus m)$, where 
$r$ is a random string generated on the fly and $\oplus$ is the bitwise 
exclusive-or operator. The algorithm $\GenAlg$, instead, simply returns a string picked uniformly at random between those whose
length is equal to that of the input. When written down as $\lamBLL$ terms, the algorithms $\EncAlg$ for encryption and $\GenAlg$ for key generation have the types in Figure~\ref{fig:privktypes}.

\subsubsection*{Defining Security}
The security of any encryption scheme, and of $\Pi_F$ in particular, is defined 
on the basis of a so-called  \emph{cryptographic experiment}, which following~\cite{katz2020introduction}
we call $\privkcpaterm^F$. Such an experiment allows the 
scheme $\Pi_F$ and a generic adversary $\Adv$ to interact. The experiment 
proceeds by first allowing $\Adv$ the possibility of generating two 
distinct messages $m_0$ and $m_1$, then encoding $m_b$ (where $b$ is picked at 
random) with a fresh key $k$, passing the obtained ciphertext $c$ to $\Adv$, and asking it to determine which one
between $m_0$ and $m_1$ the ciphertext $c$ corresponds to. The experiment
$\privkcpaterm^F$ then returns $1$ if and only if $\Adv$ succeeds in this task. In doing all 
this, the adversary is \emph{active}, i.e. it has the possibility of accessing 
an oracle for $\EncAlg_k(\cdot)$. Consequently, the 
adversary is naturally modeled as a second-order term, see again Figure~\ref{fig:privktypes}. 
Obviously, how $\Adv$ works internally is not known, but the considerations in Section~\ref{sec:polytimesoundness}
allow us to conclude that all PPT functions of that type can be encoded in 
$\lamBLL$. The security of $\Pi_F$ can be expressed as the fact that for every
such $\Adv$, the probability that 
$\privkcpaterm^F$ returns $1$ is at most $\frac{1}{2}+\varepsilon(n)$, where 
$\varepsilon$ is a negligible function. This depends, in an essential
way, on the fact that the function $F$ is indeed pseudorandom.
\begin{figure}[h]
	\centering 
	\vspace{-0.5cm}
	\begin{align*}
		\judgment{}{}{&\Gen}{\typunit \lollipop \typkeyprivk}\\
		\judgment{}{}{&\Enc}{\typkeyprivk \tensor \typmsgprivk \lollipop \typcprivk} \\
		\judgment{}{}{&\Oracle}{\typkeyprivk \lollipop \typmsgprivk \lollipop \typcprivk}\\
		\judgment{}{}{&\Adv}{\typbang{q}{(\typmsgprivk \lollipop \typcprivk)}
			\lollipop \typmsgprivk \tensor \typmsgprivk \tensor\mbox{}
			\typbang{1}{(\typcprivk \lollipop \typbool)}}\\
		\judgment{}{}{&\privkcpaterm^F}{\typbool}\\
		\judgment{}{}{&D}{!_{q}(\typstr{p_m} \lollipop \typstr{p_c}) \lollipop \typbool}
	\end{align*}
\caption{Types for Terms in the CPA-security Proof}
\label{fig:privktypes} 
\vspace{-0.5cm}
\end{figure}
\subsubsection*{Proving Security}
How is the security of $\Pi_F$ actually \emph{proved}? In fact, the proof is, like most
cryptographic proofs, done \emph{by reduction}. In other words, it proceeds contrapositively,
turning any hypothetical adversary $\Adv$ for $\Pi_F$ into a \emph{distinguisher} $D$ for $F$
(namely an algorithm designed to distinguish $F$ from a truly random function). If $D$ can be
proved successful whenever $\Adv$ is successful, we can conclude that $\Pi_F$ is secure
whenever $F$ is pseudorandom, both notions being spelled out as the \emph{non-existence}
of adversaries of the appropriate kind.

The aforementioned reduction can actually be organized as follows. First of all, we have to 
define how a distinguisher $D$ can be defined with $\Adv$ as a subroutine. The
idea is to design $D$ in such a way as to create the right environment around $\Adv$, letting
it believe that it is interacting with the experiment $\privkcpaterm$, and exploiting its capabilities for the sake
of distinguishing $F$ from a random function. In the context of $\lamBLL$, the distinguisher
$D$ becomes an ordinary term having the type in Figure~\ref{fig:privktypes}. 

Then, we have to form two instances on $D$ namely that interacting with the pseudorandom
function $F$, which we indicate as $D^F$, and that interacting with a genuinely random
function $f$, indicated as $D^f$. Both $F$ and $f$ can be assumed to be terms of $\lamBLL$,
but while the former can be taken as a term which does not use any reference, the latter can only  
be captured by a stateful computation --- one cannot hope to pick uniformly at random
a function on $n$-bit strings in polynomial time in $n$ without the help of some bookkeeping 
mechanism. The latter will actually be implemented as a reference, call it $\mathit{ledger}$, 
whose purpose is to keep track of the previous strings on which the function has been queried, so
that randomness can be generated \emph{only when needed}. Since the type of $f$ reflects the presence 
of $\mathit{ledger}$, the type of $D$ is to be updated accordingly, as we are going to describe in 
the next paragraph.

For an arbitrary type $A$ and a location context $\Xi$ whose variables do not occur in $A$, 
we define $A \cdot \Xi$ inductively as follows:
	\begin{align*}
		G \cdot \Xi &:= G  & (P \otimes Q)\cdot \Xi &:= (P \cdot \Xi ) \otimes (Q \cdot \Xi ) \\
		(\oc_p^\Theta A) \cdot \Xi  &:= \oc_p^{\Theta, \Xi} (A\cdot\Xi) &(P \arrowRef{\Theta} A)  \cdot \Xi  &:=(P\cdot\Xi) \arrowRef{\Theta, \Xi} (A\cdot\Xi)
	\end{align*}
This operation can be easily extended to term variable contexts as follows:
\[
\varnothing  \cdot \Xi  := \varnothing \qquad  (\Gamma, x : P) \cdot \Xi  := \Gamma  \cdot \Xi , x : P  \cdot \Xi 
\]
\begin{lemma}
	For every derivable judgments $\Gamma ; \Theta \vdash M : A$ and $\Gamma \vdash V : A$ and 
	every location context $\Xi$ whose variables do not occur in $\Gamma$, $\Theta$ and $A$, we obtain that
	the jugdments $\Gamma  \cdot \Xi ; \Theta, \Xi \vdash M : A \cdot \Xi $ and 
	$\Gamma \cdot \Xi  \vdash V : A \cdot \Xi $ are also derivable.
\end{lemma}
We also have to define an encryption scheme $\Pi_f$ which is structurally identical to $\Pi_F$, but which works
with truly random functions (as opposed to \emph{pseudorandom} ones) as keys. Accordingly, one can form 
a variation $\privkcpaterm^f$ on $\privkcpaterm^F$.

Now, the security of $\Pi^F$ becomes the equation $\privkcpaterm^F \coarseq\flipcoin$, whereas $\flipcoin$ is the term, of boolean type, returning each possible result with probability $\frac{1}{2}$, while $\coarseq$ is a relation coarser than $\approx$ defined by observing, through marginals, only the actual boolean value returned by the computation, without looking at the underlying store.
The aforementioned equation can be proved under the hypothesis that $F$ is pseudorandom, and this last condition also becomes an equation. This time, however, the terms to 
be compared are $D^F$ and $D^f$.

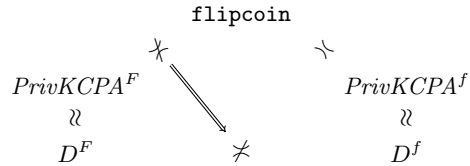
\begin{figure}[h]
	\centering
		\vspace{-3mm}
	\begin{tikzpicture}[yscale=0.9,xscale=1.1]
		\node  at (0,0) {$\flipcoin$};
		\node[rotate=45] (A) at (-1,-0.5) {$\not\coarseq$};
		\node[rotate=-45]  at (1,-0.5) {$\coarseq$};
		\node  at (-2, -1) {$\privkcpaterm^F$};
		\node  at (2, -1) {$\privkcpaterm^f$};
		\node[rotate=90]  at (-2,-1.5) {$\approx$};
		\node[rotate=90]  at (2,-1.5) {$\approx$};
		\node  at (-2, -2) {$D^F$};
		\node  at (2, -2) {$D^f$};
		\node (B) at (0,-2) {$\not\coarseq$};
		\draw[-{Implies},double] (A) to (B);
	\end{tikzpicture}
	\caption{Outline Proof of Security}\label{fig:proofscheme}
	\vspace{-1cm}
\end{figure}

The security proof then proceeds by contraposition, as explained schematically in Figure \ref{fig:proofscheme}: from the negation of the thesis, the negation of the hypothesis is derived and this is done by proving that both on the right and on the left sides of the 
diagram it is possible to \emph{link} the terms through the relation $\approx$. In this 
context, it is clear that the use of observational indistinguishability and logical 
relations becomes useful. In particular, a number of equations can be used, 
as discussed in the following section. Noticeably, all of them can be proved sound for 
observational indistinguishability through logical relations.

\subsubsection{Examples of Indistinguishability Equivalences}
Term equations and behavioral equivalences have been generalized to the setting of metric spaces via quantitative (in)equations $M =_\varepsilon N$~\cite{mardare_quantitative_2016}
and behavioral metrics~\cite{baldan2018coalgebraic,gavazzo2018quantitative}.
In our case, terms $M= \{M_n\}_n$ are families indexed by the security parameter and compared to the previous approaches, the contextual indistinguishability relation $M \sim N$ intuitively means that for every $n$, $M_n =_{\varepsilon(n)} N_n$ for some negligible function $\varepsilon$. 

We present below two typical examples of pairs of terms which are in the contextual indistinguishability relation and are used for proving security properties. A first example is the pair
\vspace{-0.1cm}
\begin{equation}
		\label{eqn:randF}
		\tag{randF}
	\return{} \false \sim \letin{y = M}{\letin{x= \random}{\equals(x,y)}}
\end{equation}
where $M$ is any computation term $\vdash M : \SS[i]$. Intuitively, it means that for a given binary string represented here by $M$, testing equality with a randomly generated string returns true with a negligible probability.

The function symbol $\random$ is interpreted as the \emph{uniform} distribution in $\D(\{0,1\}^n)$ given by  $\den{\random}_n:s\mapsto \frac{1}{2^n}$ for $n \geq 1$. The function symbol $\equals$ is interpreted as the function mapping a pair of strings $(s_1, s_2) \in \{0,1\}^n$ to $\delta_{\true}$ if $s_1 = s_2$ and to $\delta_{\false}$ otherwise for all $n \geq 1$.
Therefore, we obtain that for all $n \geq 1$, the final (big step) semantics of $\return{} \false$ is given by $\sem{\return{} \false}_n = \delta_{\false} \in \D(\{\true, \false\})$ and for $N :=\letin{y = M}{\letin{x= \random}{\equals(x,y)}}$, we have:
		\[
		\sem{N}_n = \sum\limits_{s\in \{0,1\}^n} \sem{M}_n (s) \cdot \left( \frac{1}{2^n} \delta_{\true} + \left(1 - \dfrac{1}{2^n}\right) \delta_{\false} \right)= \frac{1}{2^n} \delta_{\true} + \left(1 - \frac{1}{2^n}\right) \delta_{\false}
		\]
	We can easily see that $\return{} \false$ and $N$ are \emph{not} contextually equivalent since they reduce to different distributions. They are however contextually indistinguishable which would be quite difficult to prove directly since it requires to quantify over all closing contexts which can possibly copy their argument. 
	
	Instead, we use the logical metric defined in Section \ref{sec:logicalMet}, which here coincides with statistical distance (Definition \ref{def:statDistance}) and obtain that for all $n \geq 1$, $\dC_n^{\BB}(\return{} \false, N) = \frac{1}{2^n}$.
	Since the function $\varepsilon: n \mapsto \frac{1}{2^n}$ is negligible, the pair $(\return{}
	\false, N)$ is in $\CInd(\BB)$ and we obtain $\return{}
	\false \sim N$ by soundness (Theorem \ref{thm:soundnessContextInd}).

Another required equation states that sampling a random string is equivalent to random sampling followed by a performing a $\xor$ operation by a fixed string (represented by a computation term $\vdash M : \SS[i]$):
\begin{equation}
		\label{eqn:randXOR}
		\tag{randXOR}
		\random   \sim \letin{y = M}{\letin{x= \random }{\xor(x,y)}}
\end{equation}
The equation above is an example of \emph{Kleene equivalence} as the two terms $\random  $ and $P:= \letin{y = M}{\letin{x= \random }{\xor(x,y)}}$ have the same final (big step) semantics. The function symbol $\xor$ is interpreted by the standard \emph{exclusive-or} function on binary strings mapping a pair $(s_1, s_2)$ to $\delta_{\xor(s_1, s_2)}$. The final semantics of $\sem{P}_n$ for $n \geq 1$ is therefore given by:
\[
\sem{P}_n(s) = \sum_{s_2} \sem{M}_n(s_2)  \cdot \left( \sum_{s_1} \frac{1}{2^n} \delta_{\xor(s_1, s_2)}(s) \right) = \sum_{s_2} \sem{M}_n(s_2)  \cdot \frac{1}{2^n} = \frac{1}{2^n}
\]
where the penultimate equality holds since $\delta_{\xor(s_1, s_2)}(s) =1$ if $s = \xor(s_1, s_2)$ (or equivalently $s_1 = \xor(s, s_2)$) and $\delta_{\xor(s_1, s_2)}(s) =0$ otherwise. Since $\sem{P}_n=\sem{\random }_n$,
the distance between the two terms is therefore equal to $0$ for the logical metric and we can conclude immediately that they are in particular contextually indistinguishable. We can prove more generally that if two closed terms are Kleene equivalent, then they are contextually indistinguishable.

\section*{Related Work}
	\newcommand{\EasyCrypt}{\texttt{EasyCrypt}}
\newcommand{\Squirrel}{\texttt{Squirrel}}
Although the literature regarding formal methods for the security analysis of 
protocols and primitives is much more abundant in the symbolic model  than 
in the computational one, it certainly cannot be said that the latter has not 
been the subject of attention by the research community. The work on 
probabilistic relational Hoare logic which gave rise to the \texttt{EasyCrypt} tool~\cite{CSV2019}, 
must certainly be mentioned. The result of Bana and Comon Lundt on 
inconsistency proofs as security proofs~\cite{bana2014computationally}, 
which in turn gave rise to the \texttt{Squirrel} tool~\cite{baelde2024squirrel}, is another pertinent 
example. In both cases, the model 
provides for the possibility of higher-order constructions, which however are 
not fully-fledged. In particular, managing complexity aspects and higher-order 
functions at the same time turns out to be hard.

This last direction is the one followed by the work on CSLR and its 
formalization~\cite{nowak2010calculus}. In this case we find ourselves faced with 
a $\lambda$-calculus for polynomial time and its application to the study of cryptographic 
primitives. There are two differences with this work. First of all, the greater 
expressiveness of $\lamBLL$ allows to capture PPT even for 
second-order constructions. Furthermore, the logical relations introduced here 
effectively give rise to a notion of metric, while in CSLR the underlying 
equational theories are exact, even though a notion of observational 
equivalence similar to ours has been introduced.

Another attempt that goes in the same direction as ours is the work by Mitchell 
et al.~\cite{PPT4}, who introduced a process algebra in the style of Milner's CCS capable of 
modeling cryptographic protocols. Unlike ours, the resulting calculus is 
concurrent and this gives rise to a series of complications. Once again, 
despite the underlying notion of observational equivalence being approximate 
and therefore adhering to computational indistinguishability, the proposed 
notion of bisimulation is exact and as such much finer.

Logical relations \cite{tait1967intensional,goos_notes_1993} are a powerful tool for relational reasoning about higher-order terms. 
They are known to work well in calculi with effects and in particular in presence of
 probabilistic choice effects \cite{goubault2008logical,bizjak2015step,aguirre2023step}. It is also known that metric versions of logical relations can 
be given, and that they are useful for sensitivity analysis~\cite{reed_distance_2010,dal2022relational}. The possibility of applying logical relations to calculi such as the 
cryptographic $\lambda$-calculus is well-known~\cite{marcinkowski_complete_2004}, but the underlying calculus turns out to be fundamentally 
different from ours, being in the tradition of the symbolic model and abstracting away from probabilistic effects and 
complexity constraints.

\section*{Conclusion}
	This work shows how an approximate form of logical relation can be defined and proved sound for computational indistinguishability in a higher order $\lambda$-calculus with probabilistic effects and references. This allows cryptographic proofs to be carried out in a purely equational way by justifying the equations used.

Possible topics for future work include the transition to a logic in the style of higher-order logic, this way enabling the combination of relational and logical reasoning, in the sense of the work of Aguirre and co-authors \cite{aguirre_higher-order_2021}.

\begin{credits}
	\subsubsection{\ackname} The first two authors are partially supported by the MUR FARE project CAFFEINE,
	and by the ANR PRC project PPS (ANR-19-CE48-0014). The third author is partially supported by Fondation CFM.
\end{credits}

	\bibliographystyle{splncs04}
	\bibliography{biblio}

\end{document}